# Resilience-based Electric Sector Optimization in Response to Extreme Weather Conditions with Distributed Generation Systems

Rouzbeh Shirvani

*Department of Energy, Politecnico di Milano, Italy. E-mail: rouzbeh.shirvani@mail.polimi.it*

Tarannom Parhizkar

*The B. John Garrick Institute for The Risk Sciences, University of California, Los Angeles (UCLA), United States. E-mail: Tparhizkar@g.ucla.edu*

Extreme weather events stemming from climate change can cause significant damage and disruption to power systems. Failure to mitigate and adapt to climate change and its cascading effects can lead to short and long-term issues. The profound costs of outage in power systems, integrated with the impacts on individual safety and security from loss of critical services, necessitate an urgent need to guarantee resilience in electric power systems. This article proposes a framework to optimize the electricity sector design to be more resilient to climate change and extreme weather events by using distributed generators. The proposed framework considers components' dynamic behavior and interdependencies under an uncertain environment. The climate data and socio-economic factors are used to generate demand and supply pattern scenarios. The generated scenarios are simulated and utilized in a stochastic optimization model to find the optimal resilience-based system design.

## 1. Introduction

Climate change acceleration increases the frequency and extent of extreme weather events (IPCC Fifth Assessment Report, 2014). These events can impose threats to humans, built and natural systems. Failure to mitigate and adapt to climate change can result in short and long-term issues (Field et al., 2012). For instance, in the last ten years, distributed network failure due to heatwaves has affected one million people in Milan, Italy (Bosisio et al., 2022).

Climate change and hazardous weather events challenge all components of the power system, from generation through transmission and distribution networks to end users. The first challenge relates to electricity supply, specifically renewable energy sources. Changes in weather parameters can impact electricity supply differently. A higher frequency of drought and precipitation pattern change may unfavorably impact hydropower generation and affect cooling water availability in nuclear and thermal power plants. Coffel and Mankin (2021) analyzed the dependence of the output of thermal power plants, such as coal, nuclear, oil, and natural gas-fired plants, on ambient temperature based on the observed data of electricity output curtailments. The results anticipate a further increase in electricity output curtailments by each degree Celsius of additional global mean temperature increase. Other resources' efficiency might also be affected (Johnston et al., 2012). For example, wind generation is expected to be shut down in extremely high-speed winds to protect the rotor. The second challenge is damage caused by severe weather events to transmission and distribution networks. Heatwaves, for example, can reduce the heat transfer from underground distribution lines to soil and cause heavy thermal stresses and possible failure in the system (Bellani et al., 2022). US department of energy (2015) has reviewed the impact of weather events on grid systems:

(i) Heatwaves and high temperatures can increase transmission lines' resistance and energy loss.
(ii) High-speed winds can damage overhead distribution and transmission lines by blowing debris against lines or tower collapse.

(iii) Heavy snow and ice accumulation can gather on insulators, bridge them, and provide a conducting path, leading to a flashover fault.
(iv) A lightning strike can cause short-circuit and temporarily activate electrical protection and disconnection of the lines. Although it is generally transient and can be restored quickly, the voltage surge in the line can damage the equipment such as transformer wings.
(v) Floods and rain can damage substation equipment but not threaten overhead transmission lines.

The last challenge is increased electricity demand due to temperature increases in most regions. Climate change can influence total energy demand (particularly electricity demand) in a region directly because of changes in heating and cooling demand and indirectly due to changes in economic activity (S.Mirasgedis et al., 2007). The analysis of data on maximum daily electricity demand and temperature in Milan, Italy, shows a direct relationship between the increase in electricity demand and the ambient temperature, especially during June and July, mainly due to air conditioning purposes and a drop in energy demand in August as a result of summer vacations and decrease of economic activities (Bosisio et al., 2022).

It can be observed that extreme weather can impose direct impact, like tower collapse, or indirect impact, like heat waves and high temperature, on the operation of electrical components. Replacing, upgrading, or using more robust power system components to stand against the expected impacts of extreme weather events is a rather infeasible and unrealistic solution. A more realistic and efficient solution would be short- and long-term planning to enhance the resilience of power systems to such events (Panteli et al., 2017).

Resilience is a multi-dimensional concept, including the systems' performance before and after a severe event. The resiliency of a power system is evaluated to quantify the capability of the system to effectively and efficiently respond to extreme events (Haimes, 2009). However, such events' probabilistic and dynamic nature complex the study. As a result, it is essential to include different resilience-based dimensions and considerations during the simulation and optimization process of the electricity system with a focus on Distributed generators (DGs).

Literature in power system resilience assessment can be reviewed from generation, network, and load perspectives. From a generation perspective, Watson and Etemadi (2020) modeled electricity system resilience based on the historical hurricane footprints. The authors performed Monte Carlo-based damage modeling of grid elements based on the fragility curve of substations, transmission towers and lines, traditional thermal power plants, and solar PV plants. Yuan et al. (2016) developed a two-stage stochastic optimization to increase the system resilience by utilizing DGs. In the first stage, the location and size of the DGs are determined, while the post-restoration strategy is determined in the second stage. From a Network perspective, Ma et al. (2018) developed a two-stage stochastic mixed-integer programming for resilience enhancement of electricity distribution systems under uncertainty. The first stage determines decisions regarding component hardening and resource allocation, such as DGs and switches, while the second stage evaluates operation costs due to weather events. From a load perspective, Katal et al. (2019) studied the demand response of urban building energy, aiming to increase the power system's resilience in various extreme weather scenarios.

Studies in power system resilience assessment mainly focus on modeling the damage and its duration after an event using historical statistical data. In contrast, the effect of climate change on the frequency and intensity of extreme events in the future are neglected. In addition, previous studies mainly focused on the impact of a single severe event on the system's resilience, while the cascading effect of climate change on the energy system's performance is neglected.

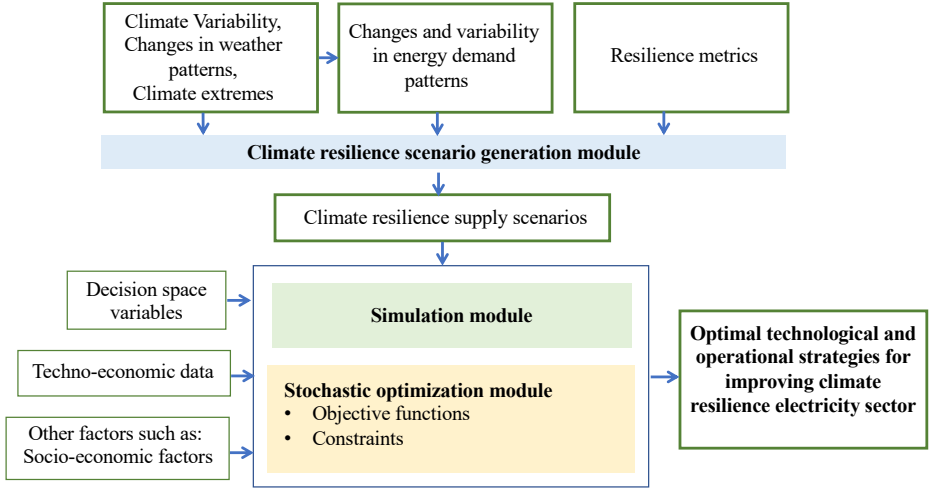

Fig. 1. Flow diagram of the proposed climate resilience-based optimization framework

This article proposes an optimization framework considering components' dynamic behavior and interdependencies in an uncertain environment. The paper is structured as follows. Section 2.1 generates demand and supply scenarios to handle the climate and socio-economic uncertainty. The effect of climate extreme events on supply and demand is then investigated to obtain climate-resilient supply and demand scenarios. In Section 2.2, the electric energy flow of distributed generation units in each scenario is simulated based on the corresponding climate variables. Section 2.3 evaluates the DGs' optimal size and operating strategies in a stochastic optimization model. Finally, the conclusion is given in Section 3.

## 2. Methodology

To enhance the resilience of the electric sector to High-Impact, Low-Probability (HILP) events and climate change, a greater understanding and capability to incorporate such events into future decision-making are needed. However, the frequency, intensity, and duration of climate extremes are changing and posing new challenges in the upcoming years. Climate forecasts can be incorporated into mitigation planning before an event happens to handle the imposed challenges. Nevertheless, climate forecasts and other socio-economic variables involve high uncertainty, including economic growth rates, demographic trends, and public policies affecting electricity demand and supply.

As a result of the involved uncertainties, other modeling methods rather than deterministic modeling is needed to consider the uncertainties of future climate and socio-economic variables. Stochastic methods as an alternative are adopted to account for the possible realization of random variables by using future scenarios.

This section reports the climate resilience-based optimization framework for the electric sector. Figure 1 presents the data flow in the proposed approach. The framework has three main modules: Climate resilience scenario generation, Simulation, and Stochastic optimization module.

The first module generates resilient demand and supply scenarios based on climate variability, climate extremes, and socio-economic factors. The generated probability-based resilience supply scenarios, along with other data, including decision space variables, techno-economic data, and other factors such as socio-economic factors, are used in the subsequent modules to simulate and optimize the system. The simulation module designs the system and energy flow of power generation units for each scenario. The stochastic optimization module evaluates optimal technological and operational strategies that improve the electric sector's resilience.

The study timeline can be defined based on the available data sets. Usually, a period of 20-30 years should be considered for the study timeline (Nik et al., 2012).

4  *Rouzbeh Shirvani and Tarannom Parhizkar.*

### 2.1. *Climate resilience scenario generation module*

The climate resilience scenario generation module utilizes climate projection datasets, climate extreme events, and socio-economic factors to generate climate resilience demand and supply scenarios. As an alternative to deal with uncertainties, explore climate change, and incorporate their effects on the electricity sector, scenario-based techniques such as Monte Carlo simulation (MCS) are implemented to generate a large set of scenarios (s). Each scenario is a possible realization of the random variables along with their probability of occurrence ($\pi_s$).

In the first step of scenario generation, the input variables' Probability Distribution Function (PDF) is obtained to generate scenarios based on the Monte-Carlo simulation. In the next step, according to the desired preciseness, distribution functions are divided into intervals with the corresponding probability. Roulette Wheel Mechanism (RWM) is then applied to generate scenarios in each time step (Niknam et al., 2012). First, the obtained probabilities are normalized to unity, and each interval is associated with an accumulated normalized probability. Next, a vector of binary parameters for each scenario is developed as follows:

$$Scenario\ s = \{B_{1,t,s}^{var_1}, \ldots, B_{l,t,s}^{var_J}\}_{t=1,\ldots,T} \quad (1)$$

Where $B \in \{0,1\}$ is the binary parameter, $J = \{1, \ldots, NV\}$ is the set of input variables, and $l = \{1, \ldots, NL\}$ is the set of intervals.

In this study, climate projection datasets obtained from Global Climate Models (GCMs) are used as input along with socio-economic variables to generate the scenarios. GCMs have been developed to forecast the future climate. However, as a result of course outputs, GCMs should be downscaled to be within the required spatial and temporal resolution. GCMs can be downscaled by employing statistical and dynamical downscaling techniques (Hellström et al., 2001). At the next step, a random number between [0,1] is generated for each input variable and time step. The first interval with an accumulated probability value less or equal to the random number is selected, the corresponding binary parameter in Eq. 1 is assigned to one, and the value of other binary parameters becomes equal to 0. The procedure is repeated until generating the desired number of scenarios. The normalized probability of each scenario can be calculated as follow:

$$\pi_s = \frac{\prod_{t=1}^{T} \prod_{J=1}^{NV} \sum_{l=1}^{NL} B_{l,t,s}^{var_J} \beta_{l,t}}{\sum_{s=1}^{NS} \prod_{t=1}^{T} \prod_{J=1}^{NV} \sum_{l=1}^{NL} B_{l,t,s}^{var_J} \beta_{l,t}} \quad (2)$$

Depending on the desired accuracy of the model, a higher number of scenarios can be generated with a more arduous computation cost. As a result, scenario-reduction methods should also be implemented to reduce the number of scenarios based on the required approximation accuracy.

The generated climate and socio-economic scenarios are transformed into electricity demand scenarios with the aid of electricity demand models. Researchers mainly used top-down and bottom-up approaches to model electricity demand in different time steps. Top-down approaches regress historical data with climate parameters. Bottom-up methods use climate parameters, such as temperature or degree-days, to estimate demand and peak load in a specific sector. There are also hybrid methods, integrating the top-down and bottom-up methods to use the advantages of each method individually (Craig et al., 2018). The demand model input variables are obtained from the developed scenarios to generate the electricity demand scenarios.

In the next step, the supply pattern can be evaluated at each time step based on the generated demand scenarios, considering the fact that demand and supply must be maintained in a constant balance, and there should always be adequate generation at each step to meet the demand. Furthermore, the generation potential of renewable energy sources is obtained from the climate variables of each scenario. The supply pattern shows the electricity supply at the system's normal performance at each time step and indicates the feasible solution area of the optimization. The performance of the electric system in the presence of extreme weather events is evaluated by vulnerability assessment of each component of the system (generation, transmission, distribution, sub-stations) based on the climate and extreme weather events parameters of the scenarios under study.

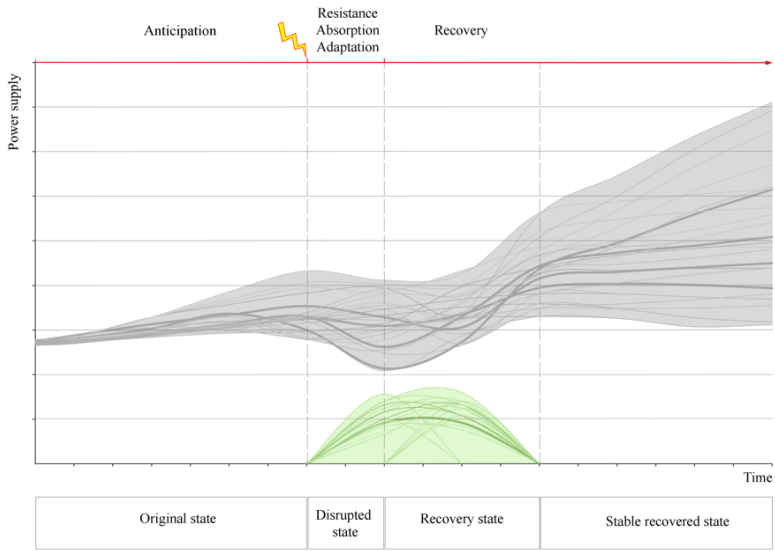

Fig. 2. Climate resilience supply scenarios over time with a disruptive event

As an example, the vulnerability of the transmission lines to wind speed can be found in the fragility curves of each transmission tower type. As a result, the probability of failure of each line at different wind speeds with the corresponding probability and the time of occurrence can be evaluated. The result shows the expected performance of the electricity system in the presence of extreme weather events. The imposed deregulations and disruptions in the power system stem from climate extremes and climate change can be compensated by multiple sets of generation units. Considering the type of generators, generation potential of renewable energy sources, the performance of the system before, during, and after the incidence of an extreme weather event, and the desired resilience level of the system to such event, multiple climate resilience supply scenarios are generated. Figure 2 shows the generated climate resilience supply scenarios.

Many resilience metrics are developed to measure the resilience of the system. The metrics evaluate the energy system's performance before and after a disruptive event without considering system-specific characteristics.

Ahmadi et al. (2021) reviewed resilience metrics considering resilience characteristics and states. The key resilience characteristics of a power system include (NIAC, 2010):

(i) Resistance: The ability of the system to resist HILP events (NIAC, 2010; McDaniels et al., 2008).
(ii) Resourcefulness: Ability of the system to manage the event effectively as it happens through available resources and assets.
(iii) Recovery and Response: The system's capacity to respond to and recover from disruptive events (NIAC, 2010).
(iv) Adaptability: Ability to boost the key resilience features before the next crisis from Lessons learned from a catastrophe (Chanda and Srivastava, 2016).

### 2.2. Simulation module

The generated climate-resilience supply scenarios are inputted into the simulation module to simulate the electric energy flow for the scenarios under study. Different software, such as PLEXOS, simulates dispatch strategies in a power system (EnergyExemplar, 2022).

The dispatch strategy determines the interactions of power generation plants with the grid, energy storage systems, and energy consumers. The strategy specifies the power generation units' operating parameters based on the generation potential of renewable energy sources, energy demand, cost of energy in the grid, and state of charge of the battery bank.

DGs are used to increase the resilience of the electric system to extreme events and tackle deregulations in power systems near energy consumers. DGs include different power sources such as Photovoltaic (PV), small Wind Turbines (WTs), Fuel Cells (FCs), diesel engines, and storage devices. As a result of the high impact of DGs on power flows, stability, and quality of power supply, in this study, DGs are considered to compensate for the disruptions that arise from climate extremes and climate change and, as a result, increase the resilience of the electric system.

### 2.3. Stochastic optimization module

The stochastic optimization module finds the optimal size of DGs, considering budget, environmental and social factors under an uncertain environment. A common approach for sequential decision-making under uncertainty is two-stage stochastic programming. In the first stage, a decision is made on parameters to find the value of random variables. In the next stage, recourse actions are made based on the obtained values of random variables. As a result of uncertainties imposed by climate projections and extreme weather events on supply and demand patterns, stochastic programming is utilized to consider model uncertainties (Perera et al., 2020).

The decision space variables of this study are the electricity generation of each technology type $i = \{WT, PV, ...\}$ at each time step $t$ and scenario $s$. As a result of the generating potential of each technology, the total electricity production by DGs in time step $t$ and scenario $s$ is presented as follows:

$$P_{DG,t,s} = [P_{WT,t,s}, P_{PV,t,s}, ..., P_{i,t,s}]_{1 \times NT_s} \quad (3)$$

Where $P_{i,t,s}$ is the power production of technology type i, in time step t and scenario s and $NT_s$ is the number of technology types in each scenario. Techno-economic data of each power generating technology is collected to map decision space variables into the objective function of the optimization.

In this study, the resilience index and cost proposed by Henry and Ramirez-Marquez (2012) are utilized to formulate the objective function. Henry and Ramirez-Marquez (2012) presented two indicators of resiliency by considering the absorption and recovery capacity of the system. It is a function of the performance of the system at a stable state before the start of disruption ($F(t_e)$), the performance of the system at the post-disruption time ($F(t_d)$), and the performance of the system at the new steady level after complete recovery ($F(t)$). The minimum value (R=0) shows the inability of the system to recover from a disruptive event, while R=1 indicates the full recovery of the system from a disruptive event.

$$R = \frac{F(t) - F(t_d)}{F(t_e) - F(t_d)} \quad (4)$$

The performance of DGs is measured based on their power supply. Total cost of the system is another indicator that is a function of the cost of implementing resilience actions ($C_{resilience\ action}$) and the cost loss of the system due to disruption ($L_{system\ disruption}$). In this study, the cost of resilience actions includes the cost of designed DGs and their operating costs.

$$C_{Total} = C_{resilience\ action} + L_{system\ disruption} \quad (5)$$

In the next step, the resilience index is mapped into the cost objective function by considering the penalty cost for the electricity needed but not served as the cost loss of the system due to disruption in the supply.

The constraints can be categorized into technical, economic, operational, reliability and resilience of the system, environmental policies, and other constraints such as demand and supply balance. In order to design DGs for the degraded power resulting from extreme events, the total electricity production is limited by demand ($P_{t,s}^{Demand}$) and the interrupted supply ($P_{t,s}^{Disturbed}$) at each time step and scenario, evaluated in the Climate resilience scenario generation module. $P_{i,t,s}$ indicates the power generation of technology $i$ at time step $t$ and scenario $s$.

$$P_{t,s}^{Disturbed} \leq \sum_i P_{i,t,s} \leq P_{t,s}^{Demand} \quad (6)$$

The generation units are limited by the generation capacity based on the social, environmental, and technological limits. As a result, the installed units are limited by minimum and maximum allowed generation capacity. $LP_i$ and $UP_i$ show the minimum and maximum generation capacity limit of each technology.

$$LP_i \leq P_{i,t,s} \leq UP_i \qquad (7)$$

The constraints may not limit those mentioned earlier, and additional constraints such as the share of specific technologies, reliability consideration of generation units, and available budget could be added to the model. Each generated supply and demand scenario is then optimized based on the minimum total cost to obtain the optimal resilient system design.

The stochastic problem is a mixed-integer nonlinear optimization problem. As a result, metaheuristic algorithms can be utilized to solve the problem under study, alleviate stagnation, and escape from local optima. Dokeroglu et al. (2019) reviewed various meta-heuristic algorithms, including new generation algorithms.

This explanatory framework will be validated for a case study in California, USA. The climate projection datasets for California state developed by W. Pierce et al. (2018) for two Representative Concentration Pathways (RCPs), RCP 4.5 and RCP 8.5, and the period between 2006 and 2050 will be used as an input to this study. The simulated wildfire by Westerling (2018) in California from 2006 to 2050 for RCP 4.5 and RCP 8.5 will be the input for climate extreme events. The vulnerability of California electricity transmission lines will be considered to study the resilience of the electric system to wildfire and high wind speeds. Techno-economic data for generating units will be obtained from IEA (2021) in three future scenarios: Stated policies, Announced pledges, and Sustainable development scenarios for 2020-2050.

## 3. Conclusion

Due to climate change acceleration, building a resilient power system and enhancing its resilience has become an indispensable requirement in power system design. Although several research studies exist in this area, it is still a new topic in power systems and needs further investigation.

While designing the electric sector based on the climate pattern changes reflects gradual changes, HILP conditions can impose significant challenges to the system. As a result, it is essential to consider climate extreme events along with climate pattern changes in long-term planning to have a more robust and resilient electric system.

Due to the uncertainty of climate changes and extreme weather forecasts, this study proposes generating different probability-based electricity supply scenarios considering different contributing factors. Simulation-based stochastic optimization of each scenario allows finding the optimal sizing and operational strategies for each scenario. By considering the different probabilities of each scenario, it is possible to find the optimal climate-resilient system designs. The developed framework can be used as a tool for researchers, agencies, and other decision-makers in the long-term planning of the electricity sector to obtain the optimum system design and operating strategy for the electric sector facing climate change and HILP events.

**Acknowledgement**

The Authors announce that no funding used for this research in its conducting process.